\documentclass[aps,prd,twocolumn,superscriptaddress,groupedaddress]{revtex4}

\usepackage{amsmath}
\usepackage{graphicx}
\usepackage{acronym}
\usepackage[colorlinks=true, allcolors=blue]{hyperref}

\acrodef{GW}[GW]{Gravitational wave}
\acrodef{LIGO}[LIGO]{Laser interferometer gravitational wave observatory}
\acrodef{MSE}[MSE]{Mean Squared Error}
\acrodef{RMSE}[RMSE]{Root Mean Squared Error}
\acrodef{FFT}[FFT]{Fast Fourier Transform}
\acrodef{SNR}[SNR]{Signal to Noise Ratio}
\acrodef{PSD}[PSD]{Power Spectral Density}
\acrodef{NF}[NF]{Normalising Flow}
\acrodef{BBH}[BBH]{Binary Black Hole}

\begin{document}

\title{Reconstructing source motion from gravitational wave strain}

\author{Joe Bayley}
\author{Chris Messenger}
\author{Graham Woan}
\affiliation{SUPA, University of Glasgow, Glasgow G12 8QQ, United Kingdom.}

\begin{abstract}
Searches for un-modelled burst gravitational wave signals return potential candidates for short duration signals. As there is no clear model for the source in these searches, one needs to understand and reconstruct the system that produced the gravitational waves. 
Here we aim to reconstruct the source motion and masses of a system based only on its gravitational wave strain. We train a normalising flow on two models including circular orbits and random unphysical motion of two point masses. 
These models are used as a toy problem to illustrate the technique. We then reconstruct a distribution of possible motions of masses that can produce the given gravitational wave signal.  
We demonstrate the ability to reconstruct the full distribution of possible mass motions from strain data across multiple gravitational wave detectors. This distribution encompasses the degenerate motions that are expected to produce identical gravitational wave strains.

\end{abstract}

\maketitle

\section{Introduction}
%
There have been \(O(100)\) \ac{GW} detections \cite{gwtc3} from compact binary systems made by the advanced ground based detectors \ac{LIGO}~\cite{advLIGOdetector} and Virgo~\cite{advVIRGOdetector}. These compact systems are well modelled using combinations of post-Newtonian theory~\cite{blanchet_gravitational_2014} and Numerical relativity \cite{NRreview, NR3D, NRtwobody}. 
Whilst compact binaries are the only type of signal that has currently been detected, other types of signals are expected within gravitational wave detectors.
A type of short duration transient signal, known as bursts do not have well defined waveform models and the signal morphology is not known before performing a search.
Searching for these signals requires un-modelled approaches that identify coherent signals between multiple detectors~\cite{cwb, bayeswave, bayeswave2}. A number of searches of this type are currently used to search for short duration ($\mathcal{O}(1)$ second or less) signals during \ac{LIGO} and Virgo observing runs~\cite{allskyburstsearch} but as yet there has been no such detection. In the event that a signal is identified as a potential astrophysical candidate, one then has to interpret the signal and identify what kind of astrophysical system produced this \ac{GW}.

In this work we propose a method for understanding the system based on finding a distribution of the possible source motions and mass distributions that could produce the observed strain from a burst detection. This could give insight into the dynamical processes of the emitting system without the need for a waveform model. 
Previous research has explored the reconstruction of density perturbations from \ac{GW} strain using classical Bayesian methods. 
In~\cite{ronaldasthesis}, motion reconstruction for simple systems is demonstrated, but in the case of binary black holes, the method struggles to accurately track how density perturbations evolve radially over time. 
Another study focuses on learning orbital dynamics using universal differential equations~\cite{keith_learning_2021}, providing point estimates of a system's dynamics from the \ac{GW} waveform. 
Additionally, several efforts have been made to reconstruct waveforms from the strain using machine learning techniques~\cite{chatterjee, denoising}, aiming to "denoise" the \ac{GW} strain and generate plausible waveforms.
In this work, however, we describe a process that directly reconstructs the dynamics or motion of masses rather than the waveform with minimal assumptions about the signal model. From these reconstructed motions, we can subsequently infer the waveform.

\begin{figure*}
\centering
\includegraphics[width=\linewidth]{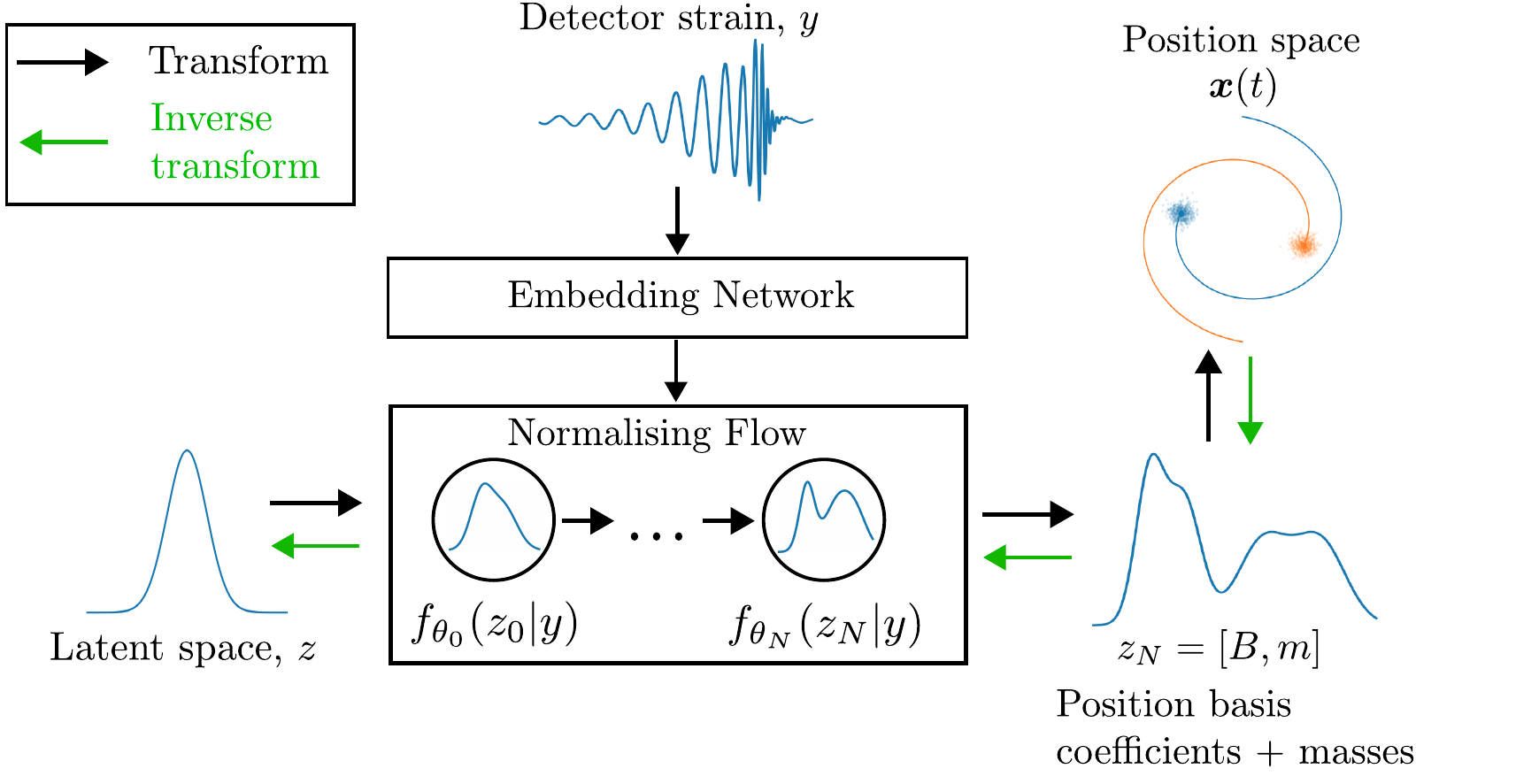}
\caption{\label{model:flowdiagram} Diagram showing how the \ac{NF} is used to invert \ac{GW} strain back to the source motions and masses. The \ac{NF} is conditioned on the detector strain data and passed through an embedding network. The output of this is then used to condition the \ac{NF}, where it learns how to transform a simple Gaussian distribution into the positional and mass space. $z$ is the latent space of the \ac{NF}, $y$ is the \ac{GW} strain from the three detectors that is input, $z_N$ is the transformed latent space after the $N$th transformation. If there are \(N\) transformations then \(z_N\) is equal to the position coefficients \(B\) and the masses \(m\). $f$ is the transformation function which is parameterised by parameters $\theta_N$. \(x(t)\) represents the positions of the masses as a function of time}
\end{figure*}

Given the inherent flexibility of machine learning techniques, they offer a promising approach to invert the \ac{GW} strain back to the system's motion (or dynamics). Our goal is to develop a method capable of reconstructing the distribution of all possible dynamics that can produce a given \ac{GW} strain, while relying on minimal prior information about the possible dynamics.
A machine learning method which is well suited to learning distributions is \ac{NF}, which provide a way to conditionally map a simple distribution to a complex target distribution, for a complete review of normalising flows see \cite{nfrev1, nfrev2}. 
\acp{NF} have been used extensively in \ac{GW} science in recent years for multiple purposes including modelling live point distribution in nested sampling~\cite{nessai, importancenessai}, generating posteriors on \ac{BBH} parameters conditioned on \ac{GW} strain \cite{dingo1, dingo2, cosmoflow, overlappingflows} and generating posteriors on cosmological parameters conditioned on \ac{GW} parameters \cite{cosmoflow}. 
In this paper we train a \ac{NF} to generate samples from a distribution representing the mass motion conditional on an input \ac{GW} time-series containing a burst signal. As a verification step we can separately compute the \ac{GW} strain for each output sample and compare to the conditional input signal. A schematic diagram showing how we use this method is shown in~Fig.~\ref{model:flowdiagram}. In this work we aim to reconstruct the the motion in simplified cases consisting of two point masses. 
Although the dynamics and \ac{GW} waveform for these systems are already well-understood, they provide a valuable test case for our technique. 
Given that inverting the strain back to the dynamics is a highly degenerate problem with numerous possible solutions, we train on an exceptionally broad and conservative prior to capture all possible dynamics, this will however maximise the degeneracy in the reconstructions. 
This paper serves as a proof of concept for our method and opens up many avenues for future development.

In Sec.~\ref{data} we describe the models used to generate the \ac{GW} strain from some masses and motions. 
The two models for circular and random motion that we use for the dynamics are described in Sec.\ref{data:models}. 
In Sec.~\ref{model} we outline the machine learning models that are used including an embedding network which interprets the raw strain data and a \ac{NF} network which learns the distribution over dynamics and masses. 
In Sec.~\ref{results} we present the resulting reconstructions of the masses and motions from all models described in Sec.~\ref{data:models} and discuss their outputs. 
The code used to generate data and train models is open source \cite{massdynamics} and the configuration files and code for the results is also available here: \cite{massdynamics_paper}.

\section{Data}
\label{data}
%
The data we use to train our \ac{NF} model consists of sets of two masses, their position as a function of time and the resulting \ac{GW} strain which results from the movement of the masses. Section~\ref{data:gwmodel} describes how we generate the \ac{GW} strain from the motion. Sec.~\ref{data:basis} describes the different choices of basis used to represent the motion and Sec.~\ref{data:models} explains the different dynamics models that we use.

\subsection{Gravitational wave model}
\label{data:gwmodel}
%
To simulate a \ac{GW} signal from any arbitrarily chosen point mass initial conditions would require the use of numerical relativity simulations which is computationally infeasible for the number of required training signals.
Therefore, we use a simplified approach to generate a waveform from an arbitrary mass motion. This simplification requires that we only consider quadrupolar \ac{GW} emission and that higher order contributions are ignored. We also constrain our analysis such that the system is made only of two point masses and that the signal always arrives from a fixed position on the sky at a fixed GPS time. The strain used throughout the paper is from three detectors, the two \ac{LIGO} detectors H1 and L1 and Virgo.

To compute the \ac{GW} strain from motions of point mass particles we follow the formalism in~\cite{flanaganhughes}.
The \ac{GW} strain tensor $h$ produced by a slowly moving source in the weak field limit can be defined in terms of the traceless mass quardupole moment $Q$ as 
\begin{equation}
    \label{data:gw:strain}
    h_{ij}(t) = \frac{2G}{c^4 d_L} \frac{d^2 Q_{ij}(t)}{dt^2},
\end{equation}
where $i$ and $j$ refer to the spatial coordinates, $G$ is the gravitational constant, $d_L$ is the luminosity distance, $c$ is the speed of light and $t$ is time. The mass quadrupole moment $I_{ij}$ is defined as 
\begin{equation}
    I_{ij}(t) = \int \rho(t, \boldsymbol{x})  x_i(t) x_j(t) d^3x,
\end{equation}
where $\rho(\boldsymbol{x})$ is the mass density as a function of position $\boldsymbol{x}$. Here we are using Cartesian coordinates where $\boldsymbol{x} = (x, y, z) \equiv (x_1, x_2, x_3)$. The traceless quadrupole moment can then be defined as
\begin{equation}
    Q_{ij}(t) = I_{ij}(t) - \frac{1}{3}\delta_{ij}I_{kk}(t),
\end{equation}
where $\delta_{ij}$ is the delta function and
\begin{equation}
    I_{kk}(t) = \sum_i I_{ii}.
\end{equation}
is the trace of the mass quadrupole moment.

Gravitational wave detectors are sensitive to the transverse and traceless part of the strain. Above we have defined the traceless \ac{GW} strain and we can follow \cite{flanaganhughes} to project the strain into the transverse (and traceless) gauge (TT gauge).
The projection tensor $P_{ij}$ defined as 
\begin{equation}
    p_{ij} = \delta_{ij} - n_i n_j,
\end{equation}
where $n_i$ is the is the $i$th component of local direction of propagation where $\boldsymbol{n} = \boldsymbol{x}/d_L$. In all models described in Sec.~\ref{data:models} the direction of propagation of the wave is assumed to be along the $z$ axis such that $n = [0,0,1]$.
We can find the transverse components of the strain by,
\begin{equation}
    h^T_{ij} = h_{kl}P_{ik}P_{jl}.
\end{equation}
We can then project $h_{kl}$ into the transverse gauge giving the transverse traceless gauge
\begin{equation}
    \label{data:gw:ttgauge}
    h^{TT}_{ij} = h_{kl}P_{ik}P_{jl} - \frac{1}{2}P_{ij}P_{kl}h_{kl}.
\end{equation}
The strain in Eq.~\ref{data:gw:strain} can then be combined with Eq.~\ref{data:gw:ttgauge} to get the final quadrupole formula
\begin{equation}
    \label{model:strain}
    h^{TT}_{ij}(t, \boldsymbol{x}) = \frac{2G}{c^4 d_L} \frac{d^2 Q_{ij}}{dt^2} \left[P_{ik}(\boldsymbol{n})P_{jl}(\boldsymbol{n}) - \frac{1}{2}P_{kl}(\boldsymbol{n})P_{ij}(\boldsymbol{n}) \right].
\end{equation}
Once we have the strain in the TT gauge, we can express it using the traditional plus and cross polarisation states as
\begin{equation}
\label{model:polarisations}
    h^{TT}_{ij} = 
    \begin{bmatrix}
        h_{+} & h_{\times}& 0 \\
        h_{\times} & -h_{+} & 0 \\
        0 & 0& 0 
    \end{bmatrix}.
\end{equation}
The strain measured at a detector combines the plus and cross polarisation terms:
\begin{equation}
    h^D(t) = h_{+}(t) F^D_{+}(t_0, \theta) + h_{\times}(t) F^D_{\times}(t_0, \theta),
\end{equation}
where \( F^D_{+, \times}(t_0, \theta) \) are the antenna pattern functions specific to detector \( D \). We use a fixed GPS time \( t_0 \) and a sky position \( \theta = (\alpha, \delta) = (\pi, \pi/2)\), where \( \alpha \) represents the right ascension and \( \delta \) represents the declination.
The strain is simulated for three ground-based detectors: the two \ac{LIGO} detectors in Hanford (H1) and Livingston (L1), and Virgo (V1). 
Here we also ignore any time delays between the detectors and assume that they are co located.

The plus and cross polarisation terms \( h_{+} \) and \( h_{\times} \), defined in Eqn.~\ref{model:polarisations}, can be explicitly written in terms of the positions \( x \), velocities \( \dot{x} \), and accelerations \( \ddot{x} \). These expressions are derived in Appendix \ref{appendixA} and shown below:
\begin{equation}
\label{model:explicitpolarisations}
\begin{split}
    h_+ &= \frac{G}{c^4 d_L} \left[ m^{(1)} \left( \ddot{x}_{1}^{(1)} x_{1}^{(1)} + \left( \dot{x}_{1}^{(1)} \right)^2 \right. \right.\\
      & \quad \quad \quad \quad \quad \quad \quad - \left. \ddot{x}_{2}^{(1)} x_{2}^{(1)} - \left( \dot{x}_{2}^{(1)} \right)^2 \right) \\
    &\quad \left. + m^{(2)} \left( \ddot{x}_{1}^{(2)} x_{1}^{(2)} + \left( \dot{x}_{1}^{(2)} \right)^2 \right. \right. \\
    &\quad \quad \quad \quad \quad \quad \quad \left. \left. - \ddot{x}_{2}^{(2)} x_{2}^{(2)} - \left( \dot{x}_{2}^{(2)} \right)^2 \right) \right],\\
    h_\times &= \frac{4G}{c^4 d_L} \left[ m^{(1)} \left( \ddot{x}_{1}^{(1)} x_{2}^{(1)} + 2 \dot{x}_{1}^{(1)} \dot{x}_{2}^{(1)} + x_{1}^{(1)} \ddot{x}_{2}^{(1)} \right) \right.\\
    &\quad \left.+ m^{(2)} \left( \ddot{x}_{1}^{(2)} x_{2}^{(2)} + 2 \dot{x}_{1}^{(2)} \dot{x}_{2}^{(2)} + x_{1}^{(2)} \ddot{x}_{2}^{(2)} \right) \right] .  
\end{split}
\end{equation}

Here, the superscript in parentheses for \( x \) denotes the mass index, while the subscript denotes the spatial dimension (e.g., \( x^{(1)}_2 \) refers to the \( y \)-dimension of mass 1). The single and double dots above \( x \) represent the first and second time derivatives, respectively. These terms are functions of time \( t \), which are omitted here for readability.
One can notice from Eqs.~\ref{model:explicitpolarisations} that there is no dependence on the \(x_3 = z\) dimension, therefore, we choose to only simulate motion in the \(x_{1,2} = (x, y)\) dimensions for simplicity. 

\subsection{Basis choice}
\label{data:basis}

We can choose to represent the mass motion \(\boldsymbol{x}(t)\) in any basis \(B\). In this work, we use two different representations of the mass motion: the time series basis and the Fourier basis.

\subsubsection{Time Series Basis}
\label{data:basis:timeseries}

In the time series basis, we parameterize the positions of the masses directly as a function of time. Here, the basis \(B\) is defined as
\begin{equation}
    B = \boldsymbol{x}(t),
\end{equation}
where \(\boldsymbol{x}(t)\) consists of four time series representing the \(x\) and \(y\) positions of each of the two masses. This basis is straightforward and allows for direct numerical computation of the time derivatives of the positions \(\boldsymbol{\dot{x}}(t)\) using second-order central differences via NumPy's gradient method \cite{numpy}. This approach is intuitive and provides a clear view of the motion over time.

\subsubsection{Fourier Basis}
\label{data:basis:fourier}

In the Fourier basis, we represent the positions as the Fourier transform of \(\boldsymbol{x}(t)\). This is defined as:
\begin{equation}
    \boldsymbol{\tilde{x}}(f) = \mathcal{F}(\boldsymbol{x}(t)) = \int_{-\infty}^{\infty} \boldsymbol{x}(t) e^{-2 \pi i ft} \, dt,
\end{equation}
where \(f\) is the frequency. The basis \(B\) is then composed of the real \(\mathcal{R}\) and imaginary \(\mathcal{I}\) components of the Fourier transform:
\begin{equation}
    B = \left[\mathcal{R}(\tilde{x}(f)), \mathcal{I}(\tilde{x}(f))\right].
\end{equation}

One advantage of using the Fourier basis is that it provides an analytical expression for the time derivatives. Specifically, the derivative in the time domain corresponds to a simple multiplication in the frequency domain:
\begin{align}
    \frac{d \boldsymbol{x}(t)}{dt} &= \frac{d}{dt} \int_{-\infty}^{\infty} \tilde{x}(f) e^{2 \pi i f t} \, df \\
    &= 2\pi i f \int_{-\infty}^{\infty} \tilde{x}(f) e^{2 \pi i f t} \, df.
\end{align}
We compute \(\tilde{x}(f)\) by taking the Fast Fourier Transform (FFT) of each positional dimension.

\subsection{Models}
\label{data:models}

There are three main dynamics models used in this analysis, these move from simpler to more complex to test the ability of the network to reconstruct mass motions. These three models are: Circular orbits with noise-free strain described in Sec.~\ref{data:models:circular}, random mass motion with noise-free strain in Sec.~\ref{data:models:rand} and random mass motion with additive detector noise strain in Sec~\ref{data:models:randnoisy}. 
For each of these models we generate training and testing data with sizes as shown in Tab.~\ref{data:normflow:table}.

\subsection{Circular orbits}
\label{data:models:circular}

We consider a system of two point masses, \( m^{(1)} \) and \( m^{(2)} \), in a circular binary orbit, simplified to lie in the \( x \)-\( y \) plane. 
In this model, the orbits do not decay due to gravitational wave emission, but we measure the quadrupolar gravitational wave strain that would be emitted from the system.

These orbits are defined over a period of 1 second with 32 samples, and can be described analytically. The relative position vector between the two objects \( \boldsymbol{x}^{(12)}(t) \) as a function of time \( t \) in two dimensions is given by

\begin{equation}
    \boldsymbol{x}^{(12)}(t) = \begin{pmatrix}
        r \cos\left( \omega t + \phi_0\right) \\
        r \cos\left(\iota \right) \sin\left( \omega t + \phi_0\right) 
    \end{pmatrix}
\end{equation}
where $r$ is the objects separation, $\phi_0$ is the initial phase, $\iota$ is the inclination and $\omega=2\pi/T$ where $T$ is the orbital period.
The orbital separation can be defined in terms of the period and masses as
\begin{equation}
    r = \left(\frac{G(m^{(1)} + m^{(2)}) T^2}{4\pi^2}\right)^{1/3},
\end{equation}
where $G$ is the gravitational constant. As the masses and positional axes are normalised as described in Sec.~\ref{data:models:circular:generation}, the value of \(G\) is not important and is set to 1. 
The positions of each object are then
\begin{equation}
\label{data:models:circular:analytic}
\begin{split}
     \boldsymbol{x}^{(1)}(t) = \frac{m^{(2)}}{m^{(1)} + m^{(2)}} \boldsymbol{r}_{12}\\
     \boldsymbol{x}^{(2)}(t) = -\frac{m^{(1)}}{m^{(1)} + m^{(2)}} \boldsymbol{r}_{12}\\
\end{split}
\end{equation}

The prior distributions for each parameter are as follows: the masses \( m^{(1)} \) and \( m^{(2)} \) are uniformly distributed in the range 0.5-1 with the constraint \( m^{(1)} > m^{(2)} \); the orbital period \( T \) is chosen such that there are between 1 and 4 cycles within the 1-second time span; the initial phase \( \phi_0 \) is uniformly distributed in the range \( 0 \) to \( 2\pi \). The inclination \( \iota \) is either \( 0 \) or \( \pi \), indicating that the orbits are either face-on or face-off.

An example of these orbits and the resulting \ac{GW} strain in three detectors can be seen in the upper panels of Fig.~\ref{data:models:positionplot}. This model differs from the random models described in Sec.~\ref{data:models:rand} as the signals are continuous rather than transient.
We choose to parameterise these signals using the time series basis defined in Sec.~\ref{data:basis:timeseries}.

\subsubsection{Data generation}
\label{data:models:circular:generation}
\begin{enumerate}

    \item {\bf Generate mass motions} Generate the masses \(m^{(1)}\) and \(m^{(2)}\) and their respective motion \(\boldsymbol{x}(t) \) using Eqs.~\ref{data:models:circular:analytic}. 
    
    \item {\bf Normalise parameters} Here, we normalise the position coefficients and the masses. The coefficients are normalised to the maximum value in the training set, which is then used as a fixed reference during testing.
    For a training set \(\boldsymbol{B}_{\rm{train}}\), the maximum value across all training data is determined as
    \begin{equation}
    B_{\text{max}} = \max \left( |\boldsymbol{B}_{\text{train}}| \right) \quad \forall \text{ training data}.
    \end{equation}
    The basis coefficients are then normalised using this maximum value with
    \begin{equation}
    B_{\text{norm}} = \frac{B}{B_{\text{max}}}.
    \end{equation}
    The masses are normalised so that their sum equals 1 for each individual data sample:
    \begin{equation}
    m^{(k)}_{\text{norm}} = \frac{m^{(k)}}{m^{(1)} + m^{(2)}},
    \end{equation}
    where \(k\) refers to the mass index.
    \item {\bf Simulate \ac{GW}} Using the positions and masses generated as described in Section~\ref{data:models:circular}, we compute the gravitational wave strain for three detectors (H1, L1, V1) following the steps outlined in Section~\ref{model}.
    
    \item {\bf Normalise strain} Each detector's strain \( h^D \) is normalised to the maximum value found across both the training set and all detectors. This ensures that the strain values are scaled consistently across the entire dataset.
    The maximum strain value over all detectors and the entire training set \(\boldsymbol{h}_{\text{train}} \) can be found with
    \begin{equation}
    h_{\text{max}} = \max \left( \boldsymbol{h}_{\text{train}}^D \right) \quad \forall D, \forall \text{ training data}
    \end{equation}
    and the strain is then normalised with 
    \begin{equation}
    h^D_{\text{norm}} = \frac{h^D}{h_{\text{max}}}
    \end{equation}
    
\end{enumerate}

\subsection{Random motion}
\label{data:models:rand}

We generate a set of random motions of masses which have not been chosen to obey any particular model of gravity and are not physically motivated. However, since physically plausible motions can be represented via the same basis, the random mass motions represent a prior volume containing physically plausible motions within it. 

For this model we choose to use the Fourier basis defined in Sec.~\ref{data:basis:fourier}.
To generate the motion we randomly sample the Fourier coefficients \(\tilde{x}(f)\) from a uniform distribution between -1 and 1.
The masses are selected uniformly within the range of 0.5 to 1, and they are normalized such that their total sum equals 1.
The scale of the basis coefficients and the strain is normalised as described in Sec.~\ref{data:models:rand:generation}.

\begin{figure}
    \centering
    \includegraphics[width=\linewidth]{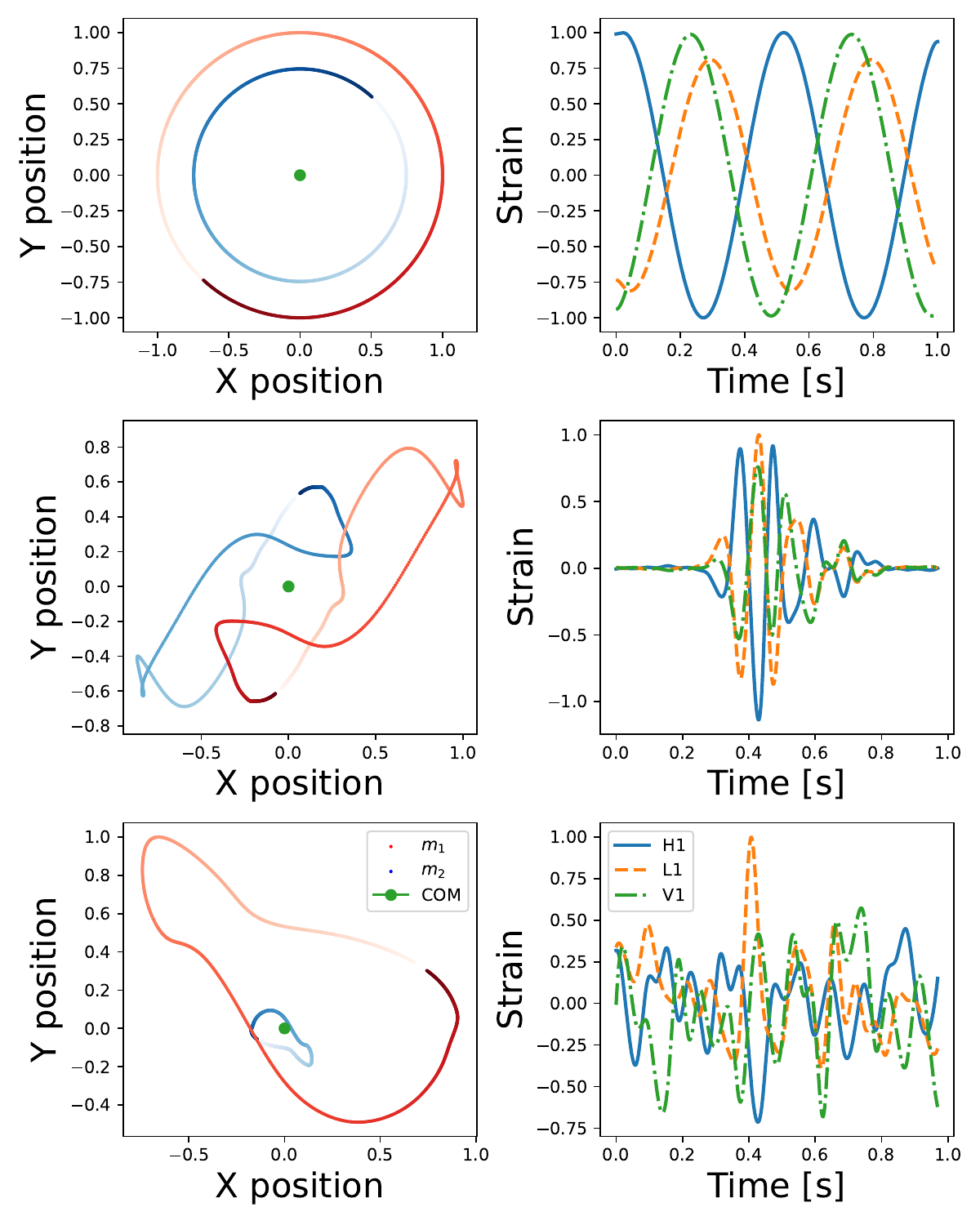}
    \caption{The positions \(x(t)\) and the strain \(h(t) \) are shown for an example of the three data models. The left panels show the motions of the two masses in the \(x\),\(y\) plane, where \(m^{(1)}\) is in blue and \(m^{(2)}\) in red, the color get darker as time progresses. The center of mass is shown as the green point. The right hand panel shows the resulting \ac{GW} strain in each of the three detectors H1 (Blue solid), L1 (orange dashed) and V1 (green dot-dash) from the motions shown in the left panel. The top row shows the circular orbit model, the lower two are the random orbit model and random orbits with noise on the \ac{GW} strain data. The transient nature of the signals in the two random models is due to the preprocessing steps applied as in Sec.~\ref{data:models:rand:generation}.}
    \label{data:models:positionplot}
\end{figure}

An example of random motions and resulting \ac{GW} strain can be seen in the second row of Fig.~\ref{data:models:positionplot}. 
The left hand panel shows the masses motion, where the symmetry is due the the fact that the motion is moved to the center of mass frame. 
We generate the strain from these motions in a three detector network (H1, L1, V1) and this is shown in the right hand panel. 
These strains are transient by design, where we window the acceleration of the motions as described in Sec.~\ref{data:models:rand:generation}. 
It is important to note that these motions are non-physical; there is no conservation of angular momentum or energy. The only constraint applied to the motion is the conservation of linear momentum, ensuring it is zero, thereby placing the system in the center-of-mass frame. The goal of this prior is to provide the network with maximum flexibility to reconstruct motion from unknown systems.

\subsubsection{Data Generation}
\label{data:models:rand:generation}

\begin{enumerate}
    \item {\bf Generate mass motions} Generate the mass motions using the desired model from Sec.~\ref{data:models}. 
    
    \item {\bf Center of Mass frame} Move motions to center of mass frame by subtracting the time varying linear momentum \(\boldsymbol{x}^{(\text{COM})}(t) \) which is found from 
    \begin{equation}
        \boldsymbol{x}^{(\text{COM})}(t) = \frac{m^{(1)} \boldsymbol{x}^{(1)}+ m^{(2)} \boldsymbol{x}^{(2)}(t)}{m^{(1)} + m^{(2)}}.
    \end{equation}
    
    \item {\bf Window acceleration} To ensure that the signals are transient, we apply a Hann window to the accelerations of the signal. First, we compute the second time derivative of the positions, \(\boldsymbol{\ddot{x}}(t)\), and then multiply it by a Hann window \(w(t)\):
    \begin{equation}
    \boldsymbol{\ddot{x}}(t) = \boldsymbol{\ddot{x}}(t) \cdot w(t).
    \end{equation}
    After windowing, we integrate the accelerations twice to return to the position space, \(\boldsymbol{x}(t)\). This process ensures that the transient signals have zero acceleration at the start and end of the time window, without altering the center of mass.

    \item {\bf Compute basis coefficients} Compute the basis coefficients \(B\) as described in Sec.~\ref{data:basis:fourier}, here we are using the Fourier domain as described in Sec.~\ref{data:basis:fourier}.
    
    \item {\bf Normalise parameters} Here, we normalise the position coefficients and the masses. This is performed in the same way as Sec.~\ref{data:models:circular:generation}. 

    \item {\bf Simulate \ac{GW}} Simulate the gravitational wave for three detectors (H1, L1, V1) using equations from Sec.~\ref{model}.
    
    \item {\bf Get strain timeseries} Compute the strain in the time domain from the basis coefficients. As we are using the Fourier domain for the basis we compute the time series using the inverse \ac{FFT} of the basis coefficients.
    
    \item {\bf Normalise strain} Each detector's strain \( h^D \) is normalised to the maximum value found across both the training set and all detectors. This is performed in the same way as Sec.~\ref{data:models:circular:generation}.

\end{enumerate}

\subsubsection{Random motion with noisy strain}
\label{data:models:randnoisy}

This dataset uses the same signal model as described in Sec.~\ref{data:models:rand}, maintaining the same priors on motion and masses, as well as the same parameterisation. The only modification is the addition of noise to the strain \( h^D(t) \). Given the arbitrary nature of the signal model, using a realistic noise power spectral density (PSD) would merely re-weight our randomly selected Fourier components, therefore we add white Gaussian noise to the strain. 
The noise \(n(t)\) is drawn from a Gaussian \(\mathcal{N}\) with fixed variance \(\sigma_n^2\).
We select \(\sigma_n\) such that is has a fixed optimal network \ac{SNR} defined as
\begin{equation}
\rho = \sqrt{\sum_D \left(\rho^{D}\right)^2},
\end{equation}
where \(\rho^D\) is the \ac{SNR} in an individual detector defined as
\begin{equation}
\rho^D = \left[ \int \frac{\tilde{h}^*(f) \tilde{h}(f)}{S_n(f)} df \right]^{1/2},
\end{equation}
 where \(h^*(f)\) is the frequency domain strain as a function of frequency \(f\) and \(S_n(f)\) is the noise \ac{PSD}.
In this example the noise is white so the noise \ac{PSD} is constant with frequency.
We choose the three detectors to have equal sensitivity with a fixed noise variance such that the optimal network \ac{SNR} \(\rho = 20\). 
The data generation follows the same steps as in Sec.~\ref{data:models:rand:generation}, however, the noise is added to the strain time series between steps 7 and 8. This is added in each detector with a fixed variance computed as described above.

\section{Model structure}
\label{model}

To recover the masses and source motions of the system, we employ a conditional \ac{NF}. This flow is conditioned on the strain time series, producing samples from the mass distribution and position coefficients. An embedding network, comprised of attention layers, is used to interpret the detector strain and reduce its dimensions before inputting it into the normalising flow. Both the embedding network and the \ac{NF} are trained simultaneously.

Training is conducted on a single Nvidia A100 GPU, with the training and testing duration's detailed in Tab.~\ref{data:normflow:table}. The AdamW \cite{adam} optimiser, with default parameters including a learning rate of \(1 \times 10^{-4}\) and a weight decay of 0.01, is used to update the network.

\subsection{Embedding network}
\label{model:embedding}

An embedding network is used to interpret the three detector strain data \(h^D(t) \)and before passing to the \ac{NF}. We use a transformer encoder, which is a set of attention layers~\cite{attention}, with varying dimensions shown in Tab.~\ref{data:normflow:table}. 
The input to the embedding network is of shape \((N_{\text{data}}, N_{\text{detectors}}, N_{\text{time-samples}}) \) the output of the attention layer is then \((N_{\text{data}}, N_{\text{embed}}, N_{\text{time-samples}}) \). After passing through \(N_{\text{attention layers}} \) the mean is taken over the time-samples dimension and this is passed through a fully connected layer of size "Linear size" in Tab.~\ref{data:normflow:table}. These embedded values \(h_{\text{embed}}\) are then passed as the condition in the \ac{NF}. 
The random and random noise models are the same size, however the circular model requires a smaller number of layers and smaller linear size due to the simpler data model.
This embedding network is trained simultaneously with the \ac{NF} described in Sec.~\ref{model:normflow}.

\subsection{Normalising flow}
\label{model:normflow}

\acp{NF} are a type of generative machine learning method that transform a distribution between a physical space \(\theta \) and a latent space \(z\) via some transformation \(f(\theta; h)\) conditioned on an observation \(h\), in our case this is the detector strain.. 
The mapping between the latent and physical space is bijective such that the transformation $z=f(\theta; h)$ has an inverse $\theta=g(z; h)$.
The probability distribution for $\theta$ can be written as
\begin{equation}
    p(\theta | h) = p(f(\theta; h)) \left| \frac{df(\theta; h)}{d\theta}\right|.
\end{equation}
Once we have an appropriate transformation we can sample from some simple distribution such as a Gaussian and transform the samples into a more complex physical space via the transformation $g(z; h)$.
The function \(f(\theta; h)\) is build of multiple transformations \(N_{\text{transforms}}\) which are constructed such that they meet the conditions that they are bijective and differentiable. 
In this work we use neural spline flows \cite{durkan_neural_2019}, where each transform is defined by a spline function, where the number of knots in the spline is defined by \(N_{\text{splines}} \) in Tab.~\ref{data:normflow:table}. 
The parameters of the splines are found by training a neural network on many examples from the data set. 
The transforms are conditioned on the output of the embedding network described in Sec.~\ref{model:embedding}. 
In our case this is the embedded \ac{GW} strain such that \(z=f(\theta; h_{\text{embed}}) \).
The physical space \(\theta\) is a combination of the basis \(B\) defined in Sec.~\ref{data:basis} and the masses \(m^{(1)}\) and \(m^{(2)}\). 
The three data models described in Sec.~\ref{data:models} require slightly different model sizes, namely the circular orbit uses less splines and transforms as the data is less complex.
We use the package {\em ZUKO} to implement \ac{NF} \cite{rozet2022zuko}.

\begin{table}[h]
\centering
\begin{tabular}{c c c c}
\multicolumn{4}{c}{\textbf{Embedding Network}} \\
& \textbf{Circular} & \textbf{Random} & \textbf{Random  Noise} \\
\hline
\(N_{\text{embed}}\)  & 32 & 32 & 32 \\
\(N_{\text{heads}}\)  & 8 & 8 & 8\\
\(N_{\text{attention layers}}\)  & 3 & 6 & 6\\
Linear size & 128 & 256 & 256 \\
\hline
\multicolumn{4}{c}{\textbf{Normalising flow}} \\
\hline
\(N_{\text{splines}}\) & 8& 16& 16\\
\(N_{\text{transforms}}\)  & 8& 16&16 \\
\(N_{\text{linear layers}}\)  & 2 & 3 & 3\\
Linear size & 128& 512& 512\\
\hline
\multicolumn{4}{c}{\textbf{Complete Network}}\\
\hline
Input size \(h^D(t) \)& (3, 32) & (3, 32) & (3, 32)\\
Output size \(\theta \)& 130& 130 & 130\\
\(N\) position coefficients & 32 & 32 & 32 \\
\hline
\multicolumn{4}{c}{\textbf{Training}}\\
\hline
N train data & \( 3 \times 10^4\)& \(5 \times 10^5\)  & \(5 \times 10^5 \)\\
N validation data & \( 3 \times 10^3\)& \(5 \times 10^4\)  & \(5 \times 10^4 \)\\
N test data & 50& 50& 50\\
Batch size & 256& 256 & 256 \\
Training Epochs & \(12 \times 10^3 \)& \( 1\times 10^4\)& \( 1\times 10^4\)\\
Training time [hours] & \(\sim 14\)& \(70\)& \(70\) \\
\hline
\end{tabular}
\caption{Model and training parameters associated with each data set. The complete network section refers to the combination of the embedding and \ac{NF} networks. The input and output size refer to the shape of the inputs \(N_{\text{detectors}}, N_{\text{timesamples}}\) and the number of predicted outputs is a combination of the 32 positions coefficients for 2 dimensions and 2 masses, plus the masses of the two objects. }
\label{data:normflow:table}
\end{table}

\subsection{Evaluating performance}
\label{model:performance}
As a measure of how well the motions are reconstructed, we can compute the \ac{RMSE} of the true strain and the strain computed from each of the reconstructed motions. In each of these cases each of the individual strain timeseries have been normalised to the absolute maximum of the true strain such that the \ac{RMSE} is computed as
\begin{equation}
    \label{results:rmse}
    \sigma^{\rm{RMSE}}_{\text{strain}} = \sqrt{\frac{1}{N}\sum_j^{N} \left(\frac{h^{\rm{recon}}(t_j) - h^{\rm{true}}(t_j)}{\max \left\{ |h^{\rm{true}}| \right\}} \right)^2},
\end{equation}
where $N$ is the number of samples in a time series and $h^{\rm{recon}}$ is the reconstructed strain and $h^{\rm{true}}$ is the true strain.

Similarly we can compute the \ac{RMSE} for the positions of the masses with 
\begin{equation}
    \label{results:rmse_pos}
    \sigma^{\rm{RMSE}}_{\text{pos}} = \sqrt{\frac{1}{N}\sum_j^{N} \left(\frac{x^{\rm{recon}}(t_j) - x^{\rm{true}}(t_j)}{\max \left\{ |x^{\rm{true}}| \right\}} \right)^2},
\end{equation}
$x^{\rm{recon}}$ is the reconstructed timeseries of mass positions and $x^{\rm{true}}$ is the true mass positions.
In Sec.~\ref{results} we compute this \ac{RMSE} on the polar coordinates rather than Cartesian coordinates, as they can more effectively highlight the degeneracies described in Sec.~\ref{degeneracies}. The polar coordinates \(r\) and \(\phi\) are defined as follows 
\begin{align}
    r &= \sqrt{x_1(t)^2 + x_2(t)^2}, \\
    \phi &= \tan^{-1}{\left(\frac{x_2(t)}{x_1(t)} \right)}.
\end{align}

\section{Expected degeneracies}
\label{degeneracies}

When observing gravitational waves, we expect to see a variety of different mass motions that can generate the same strain signal in each of the detectors. To investigate these degeneracies, we can examine the \(h_+\) and \(h_{\times}\) polarisations defined in Eqs.~\ref{model:explicitpolarisations}.
By analyzing these equations, we can identify several degeneracies that arise. For a more detailed explanation of these degeneracies, please refer to \cite{jonesdegeneracy}.

\begin{itemize}
    \item {\bf Mass symmetry} If the masses are equal $m^{(1)} = m^{(2)}$ then switching the masses will produce the same $h_{+}$ and $h_{\times}$ and the signal will appear unchanged.
    \item {\bf Reflection symmetry} If the motion is reflected across the $x$ and $y$ axes then the $h+$ and $h_{\times}$ remain unchanged. This is where all of the following transformations happen
    \begin{equation}
    \begin{aligned}
        x_1^{(n)}(t) &\rightarrow -x_1^{(n)}(t), \; x_2^{(n)}(t) \rightarrow -x_2^{(n)}(t) \\
        \dot{x}_1^{(n)}(t) &\rightarrow -\dot{x}_1^{(n)}(t), \; \dot{x}_2^{(n)}(t) \rightarrow -\dot{x}_2^{(n)}(t) \\
        \ddot{x}_1^{(n)}(t) &\rightarrow -\ddot{x}_1^{(n)}(t), \; \ddot{x}_2^{(n)}(t) \rightarrow -\ddot{x}_2^{(n)}(t),
    \end{aligned}
    \end{equation}
    where $n$ refers to either mass.
    \item {\bf Projection} As the motions are projected onto the transverse plane, any motion that is primarily along the line of sight ($z$ axis) will produce the same \ac{GW} strain. This is because there is no dependence on the $z=(x^{(n)})^3$ axis. 

\end{itemize}

\section{Results}
\label{results}

\subsection{Circular orbits}
\label{results:circular}

In the first test, we train a network on circular orbits as described in Sec.~\ref{data:models:circular}. To train the network, we generate \(3 \times 10^4\) examples of motion and compute the \ac{GW} strain for these motions across three detectors. 
To prevent overfitting, we also create a separate validation dataset consisting of \(3 \times 10^3\) examples. Finally, we use a set of 50 signals for testing. Both the testing and validation data are generated using the same priors as the training data.

\begin{figure}
    \centering
    \includegraphics[width=1.0\linewidth]{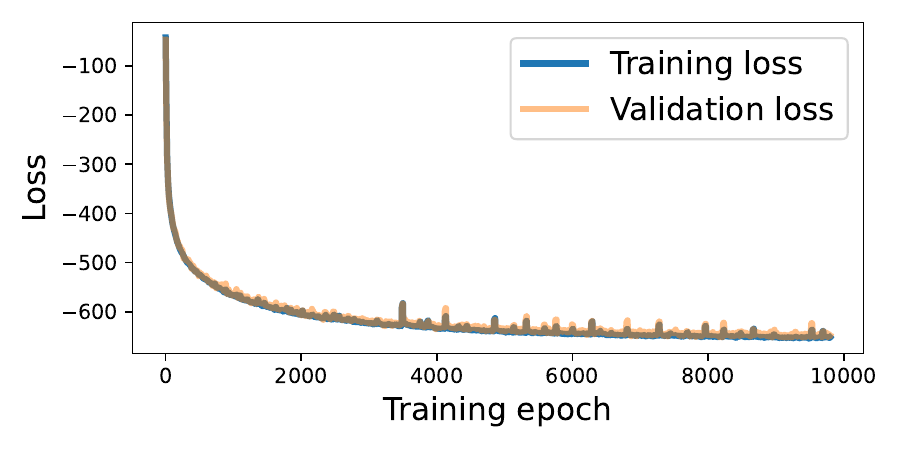}
    \caption{The training and validation losses are shown for the networks trained on circular orbits. There are $5 \times 10^5$ pieces of data in total, 90\% of these are used for training and 10\% are used for validation. }
    \label{results:circular:trainloss}
\end{figure}

The hyperparameters for the \ac{NF} and the embedding network, including the training time and number of epochs, are detailed in Tab.~\ref{data:normflow:table}. We ues the AdamW optimizer \cite{adam} and performed the training on an Nvidia A100 GPU.
The evolution of the training loss, depicted in Fig.~\ref{results:circular:trainloss}, indicates that both the training and validation curves align and converge, showing no signs of overfitting.

To test the trained network we generate a set of 50 circular orbits from the same prior as described in Sec.~\ref{data:models:circular}. 
For each of these orbits we can then compute the strain measured in each of the three detectors. 
We then input this strain into the trained network and produce a set of 500 samples from the posterior on the motions \(x(t)\) and the masses \(m^{(1)}, m^{(2)} \) for each of the 50 orbits.

\begin{figure}
    \centering
    \includegraphics[width=1.0\linewidth]{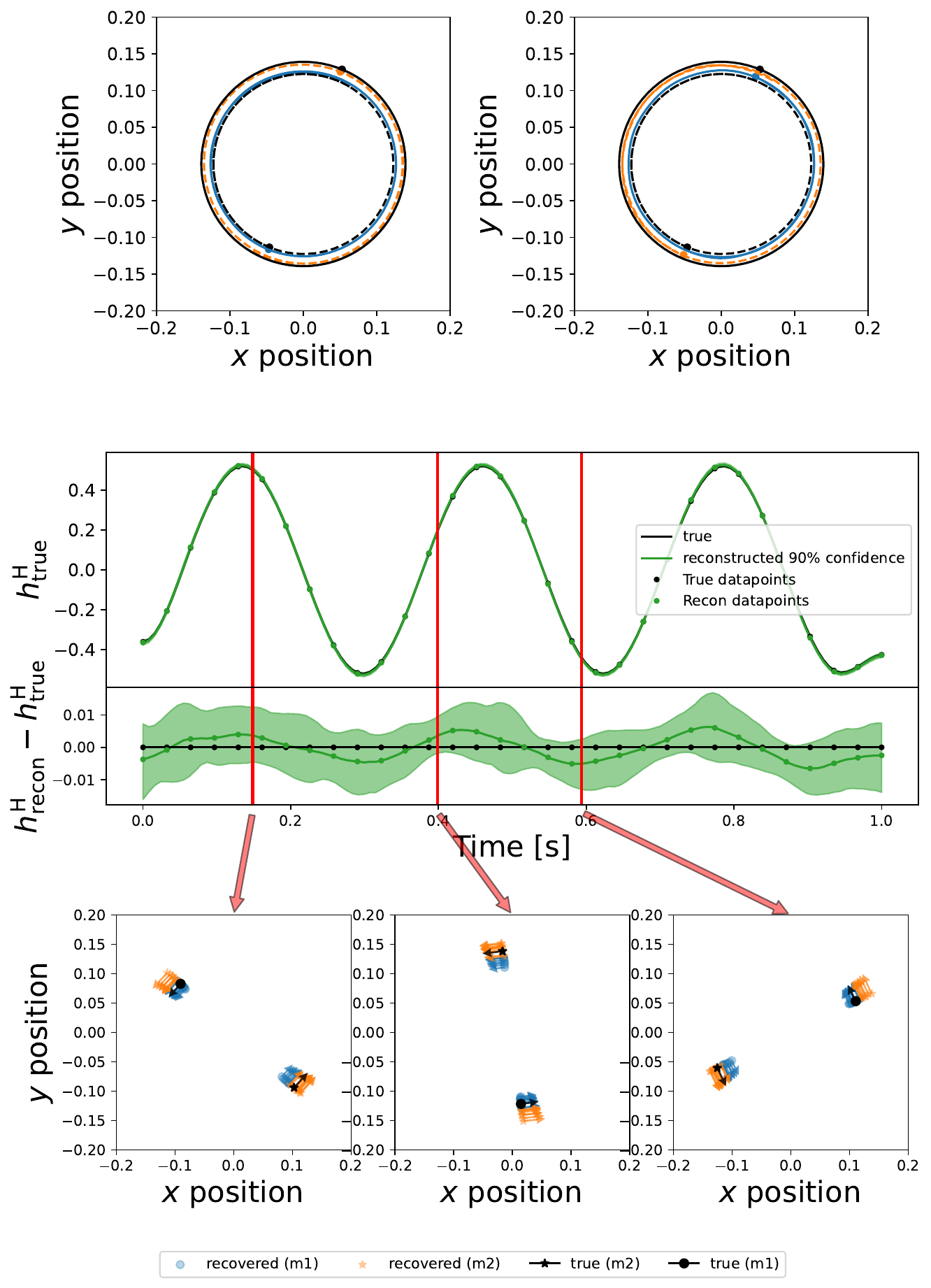}
    \caption{Here the results from a network trained on circular orbits and tested on a circular orbit are shown. The top two panels show the orbital tracks for two samples from the \ac{NF}, where dashed lines are mass 2 and solid lines are mass 1. The blue and orange colors are the reconstructed motion and the black is the simulated motion. The second panel shows the true strain (\( h_{\rm{true}}\)) in black and the median and 90\% confidence band of the strain reconstructed from the sampled motion (\( h_{\rm{recon}}\))with the true strain \( h_{\rm{true}}\) subtracted in the lower panel. In the third row, the three panels show 30 samples from posterior on \(x(t)\) at three snapshots in time referring to the three vertical red lines in the second panel. Each blue and orange pair are the reconstructed position and the black is the true injected position. The circles refer to mass 1 and the stars refer to mass 2. The arrows are the velocity vectors for each of the samples.}
    \label{results:circular:reconstruct}
\end{figure}

In Fig.~\ref{results:circular:reconstruct} we show the outputs from one of these 50 orbits. 
The top two panels show the the motion \(x(t)\) in the \(x, y\) plane for two of the 500 samples drawn from the flow and are shown alongside the true motion in black.  

\begin{figure}
    \centering
    \includegraphics[width=1.0\linewidth]{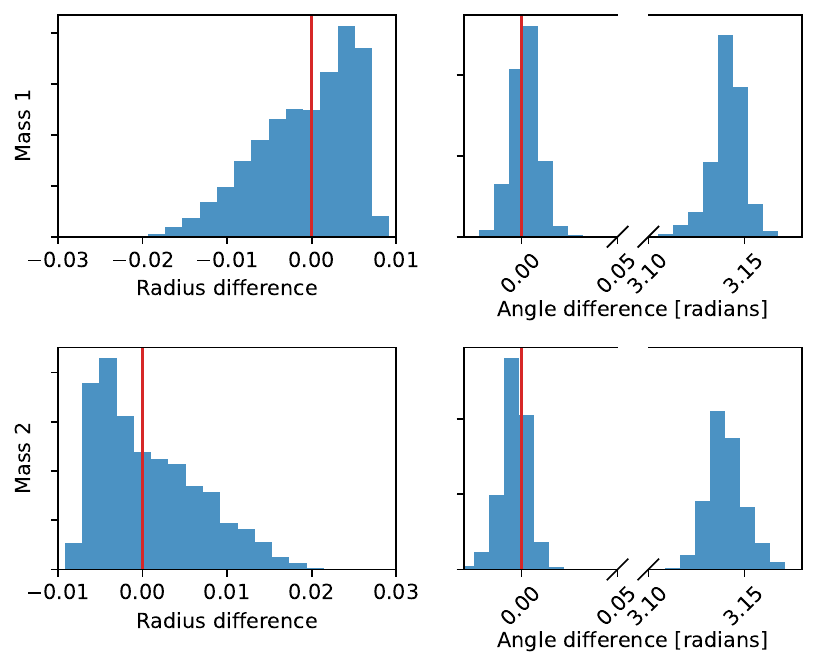}
    \caption{This shows the distribution of the differences between the true position and the reconstructed position in terms of the radius and angle of the example shown in Fig.~\ref{results:circular:reconstruct}. This distribution contains the radii or angle at every 32 time samples in all 500 reconstructed motions. The top two panels show the distribution for \(m^{(1)}\) and the lower two for \(m^{(2)}\). The left two panels show the difference between the reconstructed and true radius and the right two panels show the difference between the true and reconstructed rotational angle. The figures showing the angle distributions has been split to see the two separated distributions around 0 and \(\pi\).}
    \label{results:circular:RMSE}
\end{figure}

To assess how closely the reconstructed motion aligns with the true motion, we first convert the data into polar coordinates as described in Sec.~\ref{model:performance}. This conversion allows us to more easily identify degeneracies.
Figure \ref{results:circular:RMSE} illustrates the differences between the true and reconstructed radial \(r\) and angular \(\phi\) coordinates. The distributions include zero and are therefore consistent with the truth. The angular distribution exhibits a bimodal pattern with peaks around 0 and \(\pi\) radians, highlighting the reflection degeneracy discussed in Sec.~\ref{degeneracies}. 
This reflection corresponds to a circular orbit whose phase is shifted by \(\pi\) radians.
As detailed in Sec.~\ref{model:performance}, we compute the \ac{RMSE} between the polar radial and angular coordinates, yielding values of 0.11 and 0.48 respectively. However, when separating the two modes in the angular distribution, as shown in Fig.~\ref{results:circular:RMSE}, the \ac{RMSE} of the mode centered on zero decreases significantly to \(1.5 \times 10^{-3}\).

For each of the 500 samples drawn from the posterior on the motion \(x(t)\), we compute the \ac{GW} strain as described in Sec.~\ref{data:gwmodel}. 
For each time step of the reconstructed strains for the Hanford detector \(h^{\rm{H}}_{\rm{recon}}\), we find the 5th, 50th, and 95th percentiles and display them as the green band in the second row of Fig.~\ref{results:circular:reconstruct}. We compare this to the true strain \(h^{\rm{H}}_{\rm{true}}\), shown as the black lines. We compute the \ac{RMSE} for each of the reconstructed strains as defined in Eq.~\ref{results:rmse}, with the average being \(0.015\).

The third row of Fig.~\ref{results:circular:reconstruct} presents three snapshots of the orbits at specific time coordinates, which are indicated by the red vertical lines in the second row. These snapshots depict the positions of a subset of 30 samples from the posterior distribution on \(x(t)\) and \(m^{(1)}, m^{(2)}\).
Here the reflection symmetry described in Sec.~\ref{degeneracies} can be seen. This degeneracy results in the appearance of two distinct modes within the orange and blue samples shown in the third row of Fig.~\ref{results:circular:reconstruct}.

We can quantify how well motions are reconstructed by computing the \ac{RMSE} for the strain and polar coordinates for 50 separate orbits. Taking the average values over these 50 orbits the strain is reconstructed to an \ac{RMSE} of 0.04, the radius to an \ac{RMSE} of 0.1 and an orbital angle of 0.5 which falls to \(7.8 \times 10^{-3}\) when just taking one of the modes.

\subsection{Random motion}
\label{results:random}

In the second test, we train a network on random orbits as described in Sec.~\ref{data:models:circular}. To train the network, we generate \(5 \times 10^5\) examples of motion and compute the \ac{GW} strain for these motions across three detectors. 
To prevent overfitting, we also create a separate validation dataset consisting of \(5 \times 10^4\) examples. Finally, we use a set of 50 signals for testing. Both the testing and validation data are generated using the same priors as the training data.
We use 16 Fourier components to describe each of the motions \(x(t)\) in these tests, where each component includes real and imaginary values totalling 32 positional coefficients. 

The hyperparameters for the \ac{NF} and the embedding network, including the training time and number of epochs, are detailed in Tab.~\ref{data:normflow:table}. 
The evolution of the training loss appears similar to that for the circular orbit case and also converges with no signs of overfitting.

\begin{figure}
    \centering
    \includegraphics[width=\linewidth]{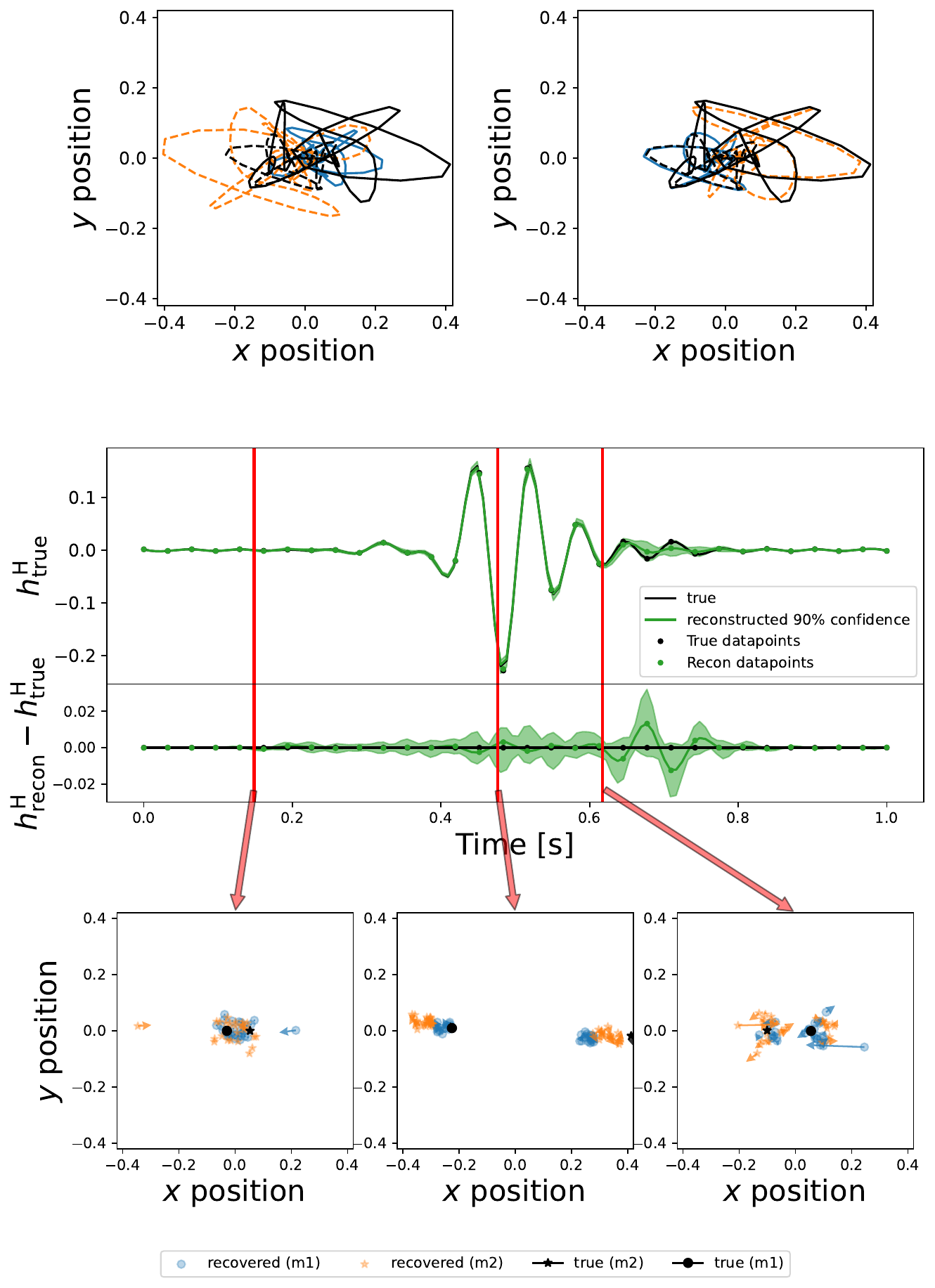}
    \caption{A model trained on random motion of masses is tested on a random motion of masses. The top row shows two examples of samples of mass motions, the labeling here is the same as used in Fig.~\ref{results:circular:reconstruct}. The second row shows the true strain (\( h_{\rm{true}}\)) in black and the 90\% confidence band on the strain reconstructed from the many mass motions (\( h_{\rm{recon}}\)) as the green band with the true strain subtracted in the lower panel. The third row shows the location and velocity vectors of 30 samples at three points in time. Here black is the truth and each blue (\(m^{(1)}\)) and orange (\(m^{(2)}\)) pair is a sample from the network. The circles represent mass 1 and stars mass 2. }
    \label{results:random:strain_pos}
\end{figure}

Once trained, we generate a set of test data containing 50 examples of motion and draw 500 samples from the posterior on the Fourier coefficients for the position and the masses of the objects. 
We compute the position as a function of time by taking the inverse \ac{FFT} of the Fourier coefficients. 
The true positions \(x(t)\) of these are shown as the black solid \(m^{(1)}\) and dotted \(m^{(2)}\) line in the first row of~Fig.~\ref{results:random:strain_pos} and the true strain that results from this is shown as the solid black line in the second row of~Fig.~\ref{results:random:strain_pos}.
We can visualise the motions from two of the samples from the network as the solid blue and orange dotted line in the two panels in the first row of Fig.~\ref{results:random:strain_pos}. 

\begin{figure}
    \centering
    \includegraphics[width=1.0\linewidth]{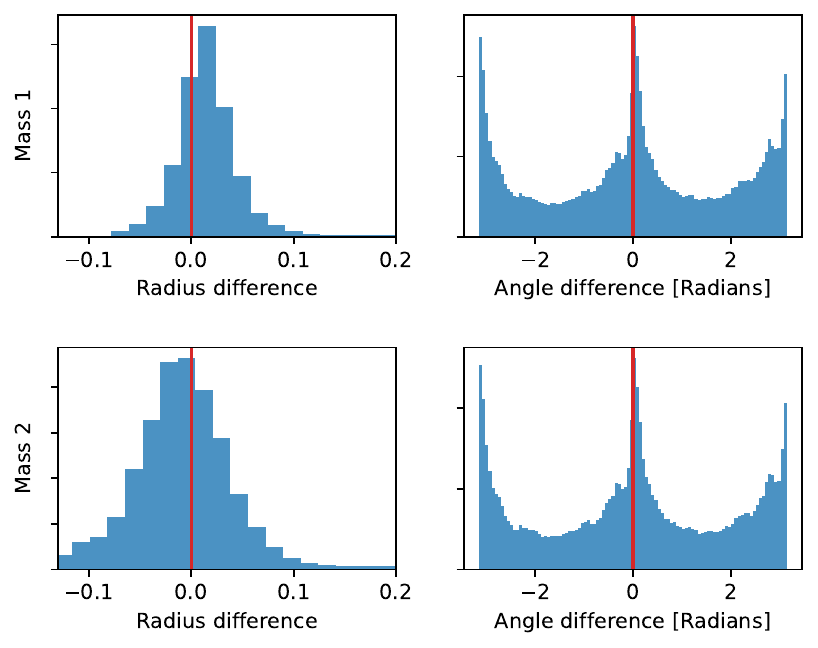}
    \caption{This shows the distribution of the differences between the true position and the reconstructed position in terms of the radius and angle of the example shown in Fig.~\ref{results:random:strain_pos}. This distribution contains the radii or angle at every 32 time samples in all 500 reconstructed motions. The top two panels show the distribution for \(m^{(1)}\) and the lower two for \(m^{(2)}\). The left two panels show the difference between the reconstructed and true radius and the right two panels show the difference between the true and reconstructed rotational angle. }
    \label{results:random:RMSE}
\end{figure}

As done in Sec.~\ref{results:circular} we can assess how closely the reconstructed motion aligns with the true motion by first converting the data into polar coordinates as described in Sec.~\ref{model:performance}. 
Figure \ref{results:random:RMSE} illustrates the differences between the true and reconstructed radial and angular coordinates. T
he distributions include and are peaked around zero, indicating successful reconstruction. 
The angular distribution exhibits a multimodal pattern with peaks around 0 and \(\pm \pi\), highlighting the reflection degeneracy discussed in Sec.~\ref{degeneracies}. 
This distribution however is broader and contains angular motion that does not match the truth well. 
This is a feature of the complex motions that that this model allows which can produce similar strains.
As detailed in Sec.~\ref{model:performance}, we compute the \ac{RMSE} between the polar radial and angular coordinates, yielding values of 0.099 and 0.62 respectively. 
Unlike the circular orbit case, it is not easy to separate out the multiple modes of the angular distribution. 

For each of the 500 samples drawn from the posterior on the motion \(x(t)\), we compute the \ac{GW} strain as described in Sec.~\ref{data:gwmodel}. 
For each time step of the reconstructed strains for the Hanford detector \(h^{\rm{H}}_{\rm{recon}}\), we find the 5th, 50th, and 95th percentiles and display them as the green band in the second row of Fig.~\ref{results:circular:reconstruct}. 
To improve visualization, the motions and strain shown in these plots have been interpolated using cubic interpolation, where they have been up sampled from 32 to 128 samples.
We only display the results for the Hanford detector, however all three are used for the \ac{RMSE} calculations. We compare this to the true strain \(h^{\rm{H}}_{\rm{true}}\), shown as the black lines. We compute the \ac{RMSE} for each of the reconstructed strains as defined in Eq.~\ref{results:rmse}, with the average being \(0.011\).

The third row of Fig.~\ref{results:circular:reconstruct} presents three snapshots of the orbits at specific time coordinates, which are indicated by the red vertical lines in the second row. These snapshots depict the positions of a subset of 30 samples from the posterior distribution on \(x(t)\) and \(m^{(1)}, m^{(2)}\).
In this case, the reflection degeneracy described in Sec.~\ref{degeneracies} can be seen. This degeneracy results in the appearance of two distinct modes within the orange and blue samples shown in the third row of Fig.~\ref{results:circular:reconstruct}.

\begin{figure}
    \centering
    \includegraphics[width=\linewidth]{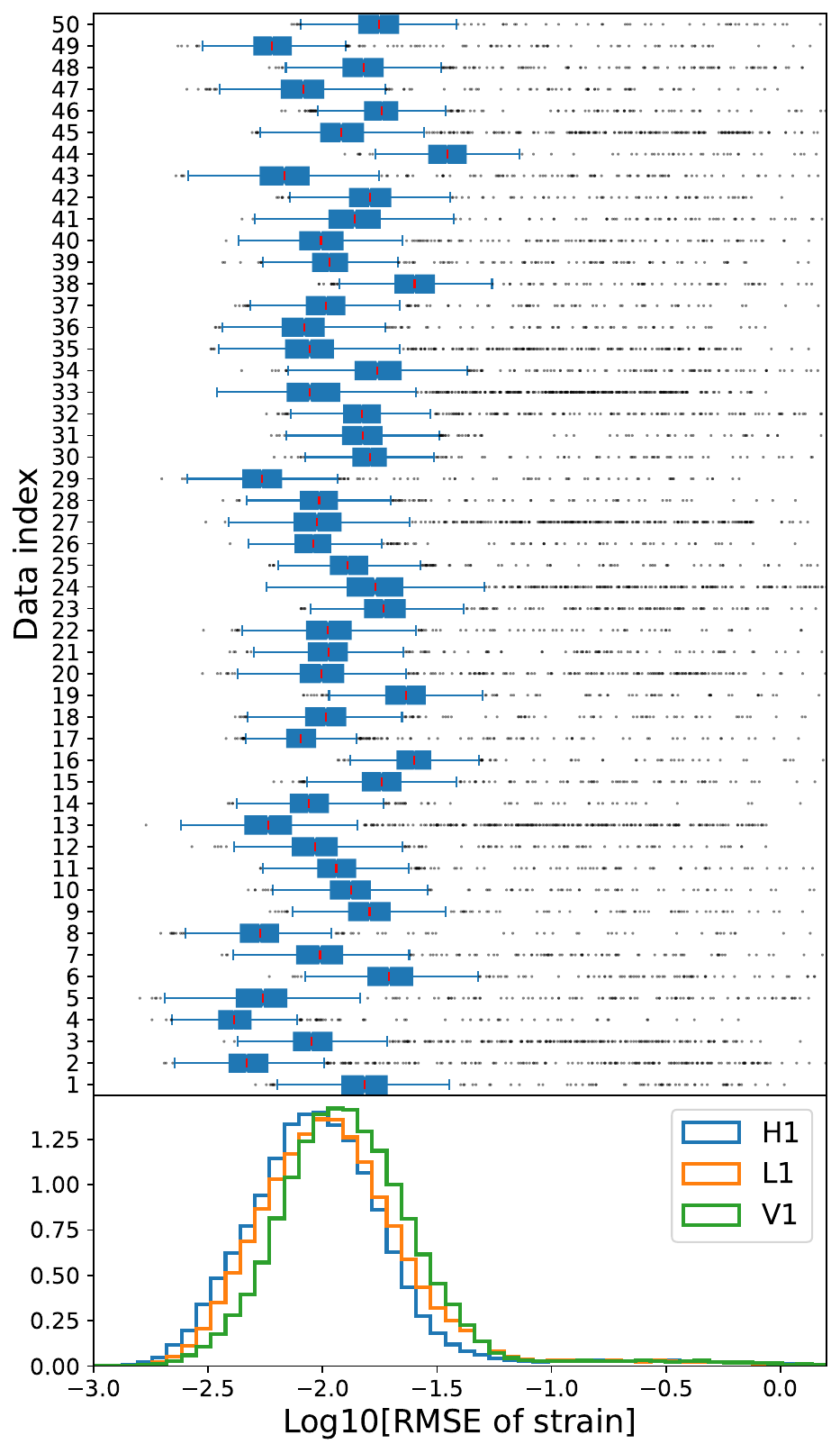}
    \caption{This figure shows the mean squared error between the true \ac{GW} strain and the recovered \ac{GW} strain from a set of samples for multiple simulations. The blue box plots in the upper panel show the distributions of the \ac{RMSE} for 50 simulations separately each containing 500 samples. These are marginalised over the three detectors.  The lower panel shown the \ac{RMSE} distribution for each detector marginalised over the set of simulations.}
    \label{results:random:mse_strain}
\end{figure}

In addition to looking at a single piece of test data, we can look at the performance of the network over a series of simulations. 
Here we choose to measure the performance of the reconstruction based on the \ac{RMSE} between the true strain and the strain computed from the motion reconstructions as defined in Eq.~\ref{results:rmse}. In Fig.~\ref{results:random:mse_strain} the distributions of \acp{RMSE} are shown for 50 test cases. In the top panel of Fig.~\ref{results:random:mse_strain} the distribution of \ac{RMSE} is shown separately for each of the 50 simulations. 
Each point in this distribution is the \ac{RMSE} for one of the 500 reconstructed strain timeseries from one of the three detectors (H1, L1, V1). The lower panel of Fig.~\ref{results:random:mse_strain} shows the distributions of \ac{RMSE} for every timestep and simulation, but computed separately for each detector. From this the median of the \acp{RMSE} over all simulations and detectors can be computed as \(0.016\).
This demonstrates that each of the motion reconstructions produces a strain close to the true simulated strain.
We also evaluate the \ac{RMSE} for the polar coordinates as in Sec.~\ref{results:circular}. The \ac{RMSE} for the radius is 0.12, while for the orbital angle it is 0.63. It is important to note that the \ac{RMSE} for the orbital angle is affected by degeneracies, which means it may not fully reflect the performance.

\begin{figure}
    \centering
    \includegraphics[width=\linewidth]{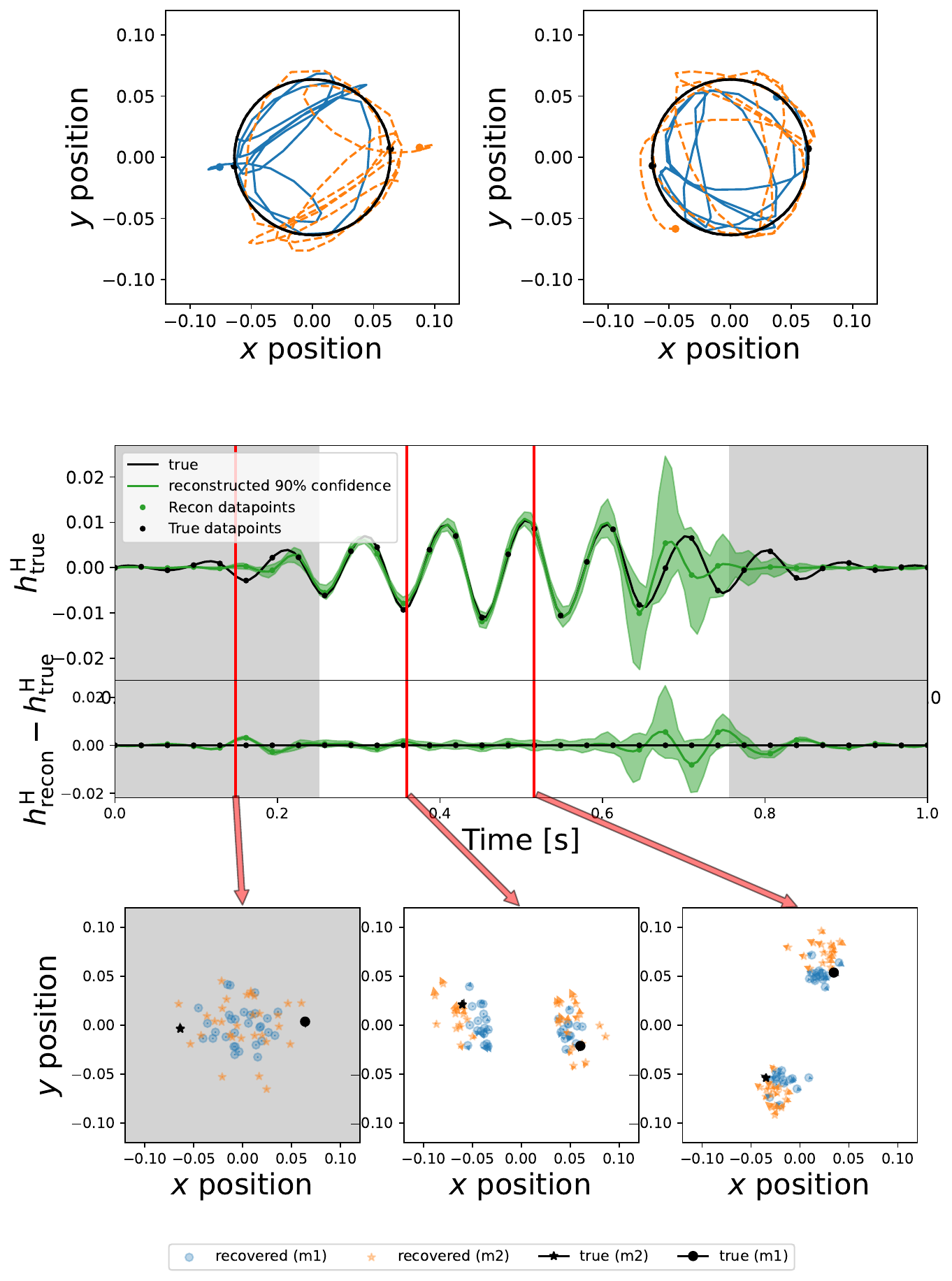}
    \caption{A model trained on random motion of masses is tested on a circular orbit model. The top row shows two samples of mass motions, the labeling here is the same as in Fig.~\ref{results:circular:reconstruct}. These tracks are truncated such that the times shown grey region in the second row is excluded. The second row shows the true strain in black and the 90\% confidence band on the strain reconstructed from the many mass motions as the green band. The third row shows the location and velocity vectors of 30 samples at three points in time. Here black is the truth and each blue and orange pair is a sample from the network. The circles represent \(m^{(1)}\) and stars \(m^{(2)}\). }
    \label{results:random:circularreconstruct}
\end{figure}

The final test for the random motion is to see how well it can recover the motion of the circular orbit model described in Sec.~\ref{data:models:circular}.
It is important to point out here that this is technically out of distribution data as the circular model is a continuous signal and the network was trained on transient windowed signals.
In the top row of Fig.~\ref{results:random:circularreconstruct} one can see the simulated circular orbit as the solid and dotted black lines and the resulting strain in the second row as the solid black strain timeseries. 
The motion \(x(t)\) in the top panel has been trimmed to only show times between times \(t_0\) and \(t_1\), the times outside this window are marked by the grey region in the second row of Fig.~\ref{results:random:circularreconstruct}.
The reconstructed motion in the top panel does not appear to closely match the truth for this reconstruction.
However, as the priors on the motion are very broad it allows all non physical motion that is consistent with the strain including circular orbits. 
It is important to point out that the motion shown in the top two two panels produces a strain consistent with the true strain despite being only partially consistent with the true motion. All of the reconstructed strains from this example give an \ac{RMSE} of 0.053.

\subsection{Random motion with noisy strain}
\label{results:randomnoise}

The noisy data uses the model described in Sec.~\ref{data:models:randnoisy}, which is the same random motion model tested in Sec.~\ref{results:random} but with noisy strain data.
The training, validation and test data sizes are the same as in Sec.~\ref{results:random} and are highlighted in Tab.~\ref{data:normflow:table}.
In training this network converged in a similar amount of time (\(\sim 70\) hours) to the random noiseless model described in Sec.~\ref{data:models:rand}. 

\begin{figure}
    \centering
    \includegraphics[width=\linewidth]{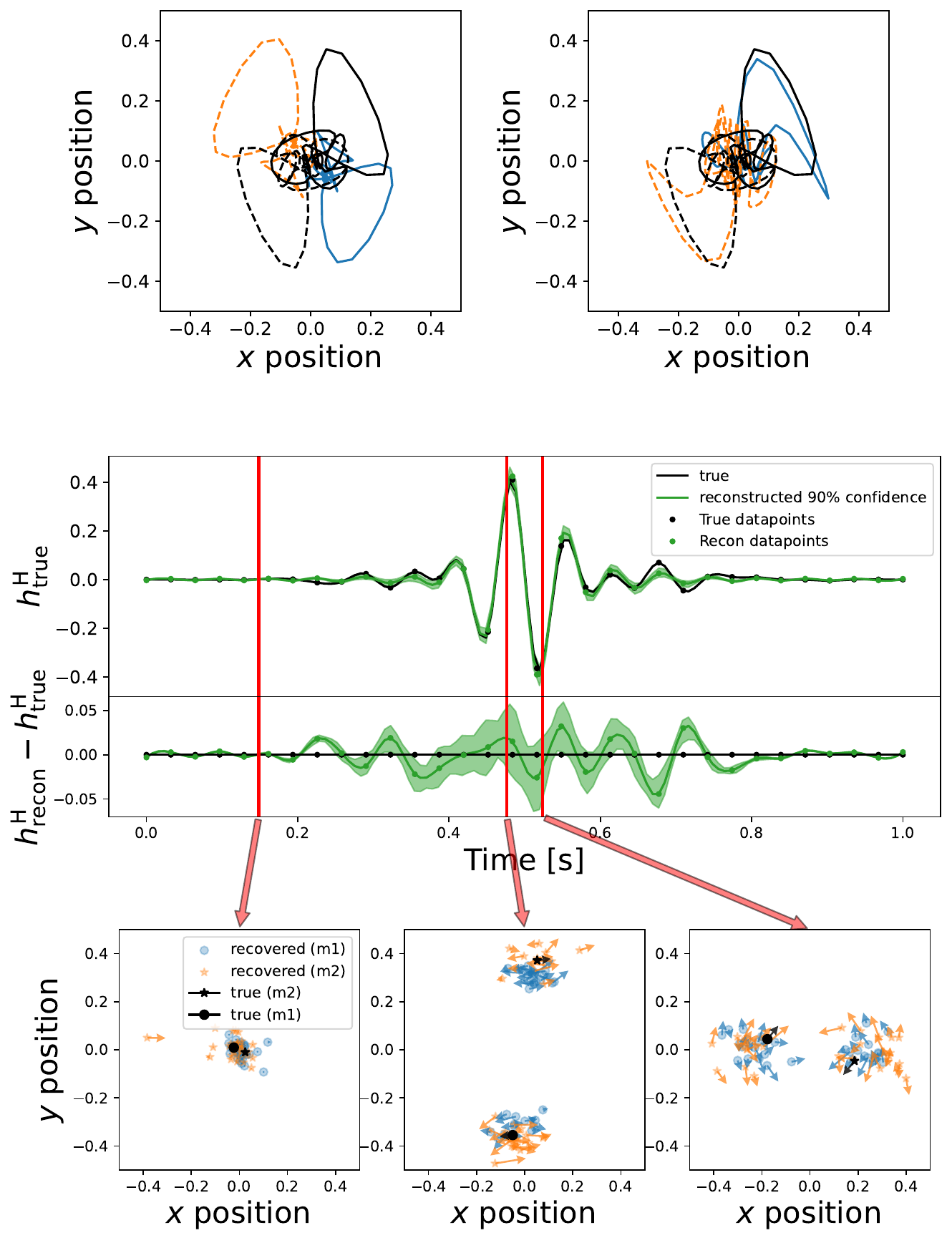}
    \caption{A model trained on random motion of masses with noisy strain data is shown. The top row shows two samples of mass motions, the labeling here is the same as in Fig.~\ref{results:circular:reconstruct}. The second row shows the true strain in black and the 90\% confidence band on the strain reconstructed from the many mass motions as the green band. The third row shows the location and velocity vectors of 30 samples at three points in time. Here black is the truth and each blue and orange pair is a sample from the network. The circles represent mass 1 and stars mass 2. }
    \label{results:randomnoise:reconstruct}
\end{figure}

In Fig.~\ref{results:randomnoise:reconstruct} we show an example of the output of a model trained on noisy strain.
We can see that the 90\% confidence band on the reconstructed strain in the second row of Fig.~\ref{results:randomnoise:reconstruct} is broader than in its noiseless counterpart in Fig.~\ref{results:random:strain_pos}. 
The \ac{RMSE} between the reconstructed and true strain is 0.013.
We can also compute the \ac{RMSE} for the distribution of motions as in Sec.~\ref{results:random}, where this is computed over the polar coordinated \((r, \phi)\).
The \ac{RMSE} for \(r\) is found to be 0.02 and for \(\phi\) is 0.67.
These values are larger than in the noiseless counterpart, however, the added noise will increase the size of the prior distribution, we would expect a larger number of degenerate motions that produce similar waveforms.

\section{Discussion}
\label{discussion}

In this work we describe a method to extract the source motion of a system from its \ac{GW} strain, with the goal of learning the structure of a system of masses that produced a \ac{GW} from an unknown origin. We achieved this by using a \ac{NF} conditioned on the \ac{GW} strain to produce samples of the motions and masses of a system.

In Sec~\ref{data} we described the data generation procedure, defining how we generate the motion of masses and how the \ac{GW} strain is computed. We also discussed the constraints that are applied to this model including enforcing that it is made of only two point masses in the center of mass frame and the \ac{GW} emission is quadrupolar.

In Sec.~\ref{model} we describe the type of machine learning method we used for reconstructions. This consisted of two parts, an embedding network and a \ac{NF}. 
The embedding network took the time series strain from three ground based detectors as input and encoded this into a vector of either 128 or 256 numbers depending on the data model. These embeddings are then input to a \ac{NF} which is trained to produce a posterior distribution of the objects masses and motion.

This method was then trained and tested on two different models: a set of circular orbits in Sec.~\ref{results:circular}, and a set of random motions of masses in Sec.~\ref{results:random}.
Both of these models were trained on $3 \times 10^4$ examples of data for 14 hours on an Nvidia A100 GPU.
We demonstrated that the \ac{NF} can reconstruct motion which either follows the true injected motion or some degenerate motion which would produce the same strain. We showed that one symmetry appears where the positions, velocities and accelerations can be inverted but still return the same strain. This aligned with the expected degeneracy from the quadrupolar waveform computation described in Sec.~\ref{degeneracies}. We found a median \ac{RMSE} between the true strain and the reconstructed strain to be $0.04$ over each of the three detectors. 

The second test was on random motions of masses, the goal here was to keep the prior distribution of motion as broad as possible. 
Within the parameter space defined by this broad set of random motions there exists the smaller subset of motions consistent with both Newtonian and general relativity. However, being this broad of a prior we can expect to reproduce many other motions which are not physically possible. 
This network was trained on a set of $5 \times 10^5$ strain timeseries and associated motions and took $\sim $ 70 hours to train on an Nvidia A100 GPU.
We have shown an example in Fig.~\ref{results:random:strain_pos} where the strain reconstructions given an \ac{RMSE} of 0.011. Over a set of 50 simulations, the median \ac{RMSE} between the true strain and the strain computed from reconstructed motion was 0.016.
We found that whilst all of the motions that are generated produce a strain similar to the truth, many of the motions exhibit random non central accelerations. 

In this work, we use a simplified model for both the waveform and mass motion. However, our future research will focus on increasing this complexity. Specifically, we plan to advance to higher-order multipole moments beyond the quadrupole and explore spherical harmonic bases for mass distributions. Separately, we aim to model more intricate systems by incorporating a larger number of point masses.
We also intent to apply more constraints on the motions of the masses to minimise the number of non physical outputs.
We also intend to extend our approach beyond the current 16 frequency components. This will involve restructuring the normalizing flow (NF) to generate samples recurrently, thereby reducing the need to estimate numerous components. We will also apply novel equivariant machine learning techniques to leverage problem symmetries effectively.
Furthermore, we plan to refine the prior space to better align with Newtonian and general relativistic models. Finally, we aim to apply this methodology to compact binary coalescence (CBC) events with broad, weak priors, enabling the generation of unparameterised mass motion conditioned on observed data.

\section{Acknowledgements}
This research is supported by the Science and Technology Facilities Council., J.B.\, G.W.\ and C.M.\ are supported by the Science and Technology Research Council (grant No. ST/V005634/1).
The authors are grateful for computational resources provided by the LIGO Laboratory supported by National Science Foundation Grants PHY-0757058 and PHY-0823459.

\bibliography{refs}

\appendix

\section{Degeneracy computation}
\label{appendixA}

We can write out the equations in Sec.~\ref{degeneracies} for the toy problem of two point masses as below.
Starting with the quadrupole formula which becomes a sum over the point masses, here we are looking at the trace reduced part of the formula.
\begin{equation}
    \tilde{Q}_{ij}(t) = I_{ij}(t) - \frac{1}{3} \delta_{ij} I_{kk}(t)    
\end{equation}
where
\begin{equation}
    \begin{split}
        I_{kk}(t) &= m^{(1)} \left( (x_1^1(t))^2 + (x_1^2(t))^2 + (x_1^3(t))^2 \right) \\ &+ m^{(2)} \left( (x_2^1(t))^2 + (x_2^2(t))^2 + (x_2^3(t))^2 \right),
    \end{split}
\end{equation}
and the $i,j,k$ indices refer to the spatial $x, y, z$ coordinates.
the first time derivative becomes
\begin{equation}
    \begin{split}
        \dot{Q}_{ij}(t) &= m^{(1)} \left( \dot{x}_1^i(t) x_1^j(t) + x_1^i(t) \dot{x}_1^j(t) \right) \\ 
        &+ m^{(2)} \left( \dot{x}_2^i(t) x_2^j(t) + x_2^i(t) \dot{x}_2^j(t) \right) \\
        &- \frac{1}{3} \delta_{ij} \left( 2m^{(1)} \sum_{k} x_1^k(t) \dot{x}_1^k(t) + 2m^{(2)} \sum_{k} x_2^k(t) \dot{x}_2^k(t) \right)
    \end{split}
\end{equation}

The second time derivative is then
\begin{equation}
\begin{split}
\ddot{Q}_{ij}(t) &= m^{(1)} \left( \ddot{x}_1^i(t) x_1^j(t) + 2 \dot{x}_1^i(t) \dot{x}_1^j(t) + x_1^i(t) \ddot{x}_1^j(t) \right) \\
&+ m^{(2)} \left( \ddot{x}_2^i(t) x_2^j(t) + 2 \dot{x}_2^i(t) \dot{x}_2^j(t) + x_2^i(t) \ddot{x}_2^j(t) \right) \\
&- \frac{2}{3} \delta_{ij} \left( m^{(1)} \sum_{k} (\dot{x}_1^k(t) \dot{x}_1^k(t) + x_1^k(t) \ddot{x}_1^k(t))\right. \\
&\left.+ m^{(2)} \sum_{k} (\dot{x}_2^k(t) \dot{x}_2^k(t) + x_2^k(t) \ddot{x}_2^k(t)) \right)
\end{split}
\end{equation}

From Eq.\ref{} if we take a plane wave travelling in the z direction we can compute the projection operator as 
\begin{equation}
    P_{ij} = \delta_{ij} - \delta_{i3}\delta_{j3}
\end{equation}

We can then apply this to the quadrupole tensor
\begin{equation}
    \ddot{Q}_{ij}^{\text{TT}} = \left( P_{ik} P_{jl} - \frac{1}{2} P_{ij} P_{kl} \right) \ddot{Q}_{kl}
\end{equation}

The first term in the above can be written as 
\begin{equation}
    \begin{split}
    P_{ik} P_{jl} \ddot{Q}_{kl} &= (\delta_{ik} - \delta_{i3} \delta_{k3})(\delta_{jl} - \delta_{j3} \delta_{l3}) \ddot{Q}_{kl} \\
    &= (\delta_{ik} \delta_{jl} - \delta_{ik} \delta_{j3} \delta_{l3} - \delta_{i3} \delta_{k3} \delta_{jl} + \delta_{i3} \delta_{k3} \delta_{j3} \delta_{l3}) \ddot{Q}_{kl}
    \end{split}
\end{equation}

The second term becomes

\begin{equation}
    \begin{split}
        \frac{1}{2} P_{ij} P_{kl} \ddot{Q}_{kl} &= \frac{1}{2} (\delta_{ij} - \delta_{i3} \delta_{j3})(\delta_{kl} - \delta_{k3} \delta_{l3}) \ddot{Q}_{kl} \\
        &= \frac{1}{2} (\delta_{ij} \delta_{kl} - \delta_{ij} \delta_{k3} \delta_{l3} - \delta_{i3} \delta_{j3} \delta_{kl} \\
        &+ \delta_{i3} \delta_{j3} \delta_{k3} \delta_{l3}) \ddot{Q}_{kl}
    \end{split}
\end{equation}

We can then combine the terms to make

\begin{equation}
    \begin{split}
        \ddot{Q}_{ij}^{\text{TT}} &= \ddot{Q}_{ij} - \delta_{i3} \ddot{Q}_{3j} - \delta_{j3} \ddot{Q}_{i3} + \delta_{i3} \delta_{j3} \ddot{Q}_{33} \\
        &- \frac{1}{2} \left( \delta_{ij} \ddot{Q}_{kk} - \delta_{ij} \ddot{Q}_{33} - \delta_{i3} \delta_{j3} \ddot{Q}_{kk} + \delta_{i3} \delta_{j3} \ddot{Q}_{33} \right)
    \end{split}
\end{equation}

As seen in Eq.~\ref{} we can use the $\ddot{Q}_{11}^{\text{TT}}$ and $\ddot{Q}_{12}^{\text{TT}}$ terms for find $h_{+}$ and $h_{\times}$, giving
\begin{equation}
    \begin{aligned}
        h_+ &= \frac{G}{2c^4 d_L} \left[ m^{(1)} \left( \ddot{x}_1^1 x_1^1 + 2 \dot{x}_1^1 \dot{x}_1^1 + x_1^1 \ddot{x}_1^1 \right) \right.\\
        &\quad + m^{(2)} \left( \ddot{x}_2^1 x_2^1 + 2 \dot{x}_2^1 \dot{x}_2^1 + x_2^1 \ddot{x}_2^1 \right)  \\
        &\quad  - m^{(1)} \left( \ddot{x}_1^2 x_1^2 + 2 \dot{x}_1^2 \dot{x}_1^2 + x_1^2 \ddot{x}_1^2 \right) \\
        &\quad \left.- m^{(2)} \left( \ddot{x}_2^2 x_2^2 + 2 \dot{x}_2^2 \dot{x}_2^2 + x_2^2 \ddot{x}_2^2 \right) \right] \\
        &= \frac{G}{c^4 D} \left[ m^{(1)} \left( \ddot{x}_1^1 x_1^1 + \left(\dot{x}_1^1 \right)^2 - \ddot{x}_1^2 x_1^2 - \left( \dot{x}_1^2 \right)^2 \right) \right.\\
        &\quad \left. + m^{(2)} \left( \ddot{x}_2^1 x_2^1 + \left(\dot{x}_2^1\right)^2 - \ddot{x}_2^2 x_2^2 - \left( \dot{x}_2^2 \right)^2\right) \right].
    \end{aligned}
\end{equation}

\begin{equation}
    \begin{aligned}
    h_\times &= \frac{4G}{c^4 d_L} \left[ m^{(1)} \left( \ddot{x}_1^1 x_1^2 + 2 \dot{x}_1^1 \dot{x}_1^2 + x_1^1 \ddot{x}_1^2 \right) \right.\\
    &\quad \left.+ m^{(2)} \left( \ddot{x}_2^1 x_2^2 + 2 \dot{x}_2^1 \dot{x}_2^2 + x_2^1 \ddot{x}_2^2 \right) \right].
    \end{aligned}
\end{equation}

\end{document}